\documentclass[%
a4paper,
5p,
twocolumn,
authoryear,
times,
fleqn,
sort&compress,
]%
{elsarticle}

\usepackage{amsmath}
\usepackage{amsfonts}
\usepackage{natbib}
\usepackage{aas_macros}
\usepackage{setspace}
\usepackage{multicol}

\usepackage{graphicx}
\usepackage{bm}
\usepackage{ifthen}
\usepackage{mathrsfs}
\usepackage{color} 

\renewcommand*{\O}{\ensuremath{\mathcal{O}}}

\renewcommand{\vec}[1]{\ensuremath{\boldsymbol{#1}}}

\newcommand*{\ten}[1]{\ensuremath{\boldsymbol{\mathsf{#1}}}}

\newcommand*{\mi}[1]{\ensuremath{\boldsymbol{\mathsf{#1}}}}

\newcommand*{\figref}[1]{Fig.~\ref{#1}}
\newcommand*{\eqnref}[1]{\ifthenelse{\equal{#1}{}}{\note{EQUATION}}{Eq.~\ref{#1}}}
\newcommand*{\eqnrefp}[1]{(Eq.~\ref{#1})}
\newcommand*{\secref}[1]{Sec.~\ref{#1}}
\newcommand*{\tabref}[1]{Table~\ref{#1}}

\newcommand*{\D}{\ensuremath{\mathscr{D}}}
\newcommand*{\T}{\ensuremath{\mathscr{T}}}
\newcommand*{\tl}[1]{\ensuremath{\overline{#1}}}

\newcommand*{\op}[1]{\textsc #1}
\newcommand*{\code}[1]{\mbox{\textsc{#1}}}

\newgeometry{margin=1.5cm,bottom=1.6cm,columnsep=0.5cm}
\begin{document}
    \singlespacing

\title{An Optimizing Symbolic Algebra Approach for \\ Generating Fast Multipole Method Operators}

\author[1]{Jonathan P. Coles\corref{cor1}}%
\address[1]{Physik-Department T38, Technische Universit\"at M\"unchen, James-Franck-Str.\ 1, D-85748 Garching, Germany}
\ead{jonathan.coles@tum.de}

\author[2]{Rebekka Bieri}%
\address[2]{Max Planck Institute for Astrophysics, Karl-Schwarzschild-Str.\ 1, D-85741 Garching, Germany}
\ead{bieri@mpa-garching.mpg.de}

\cortext[cor1]{Corresponding author}
\date{}

\journal{Computer Physics Communications}

\begin{abstract}
We have developed a symbolic algebra approach to automatically produce, verify, and optimize computer code for the Fast Multipole Method (FMM) operators.
This approach allows for flexibility in choosing a basis set and kernel, and can generate computer code for any expansion order in multiple languages.
The procedure is implemented in the publicly available Python program \code{Mosaic}.
Optimizations performed at the symbolic level through algebraic manipulations significantly reduce the number of mathematical operations compared with a straightforward implementation of the equations.
We find that the optimizer is able to eliminate 20-80\% of the floating-point operations and for the expansion orders $p \le 10$ it changes the observed scaling properties.
We present our approach using three variants of the operators with the Cartesian basis set for the harmonic potential kernel $1/r$, including the use of totally symmetric and traceless multipole tensors.
\end{abstract}


\maketitle



\section{\label{sec:Introduction} Introduction}
Many physical systems can be described by a potential function $\Phi \sim 1/r$ that scales inversely as a function of distance.
For systems modeled as a collection of $N$ discrete particles, this function acts between pairs of particles, resulting in $\O(N^2)$ interactions.
An exact calculation of $\Phi$ quickly becomes prohibitive on modern hardware even for modestly sized problems of several thousand particles.
Furthermore, following the time evolution of the particles requires $\nabla \Phi$ to be computed thousands if not millions of times.
In the fields of molecular dynamics and astrophysics many approximate methods have been developed to make their respective problems tractable.
Popular $\O(N\log N)$ methods include the Barnes-Hut tree code \citep{1986Natur.324..446B} and Particle Mesh Ewald \citep[PME,][]{1993JChPh..9810089D}, a spectral method that implicitly accounts for periodic boundary conditions. 

In recent decades, the $\O(N)$ Fast Multipole Method (FMM) \citep{1987JCoPh..73..325G} has been attracting more attention, not just because of the theoretical performance, but also because parallel implementations require less network communication; a feature that will become important in the era of exascale machines \citep{Yokota:2012:TSF:2403996.2403998}. Similar to the Barnes-Hut approach, FMM approximates sub-regions (or cells) of the simulation volume by multipole expansions of the local point sources. However, it achieves $\O(N)$ scaling by allowing cell-cell interactions, rather than only particle-cell interactions.
The accuracy is controlled by the order of the multipole expansion and by a ``well-separated'' criteria for the cells.

While the theoretical scaling of FMM is clearly computationally attractive and the multipole expansion approach is simple to understand, developing an efficient FMM implementation still requires a significant investment of time and understanding.
There is no single FMM algorithm and the specifics often depend on the problem at hand.
One must also carefully choose the basis set to represent the multipole tensors---e.g., spherical harmonics \citep{1987JCoPh..73..325G,2014ComAC...1....1D}, Cartesian \citep{2002JCoPh.179...27D}, or planewaves \citep{greengard_rokhlin_1997}---as well as the type of kernel representing the interaction potential, e.g., Laplacian, Helmholtz \citep{Greengard:1998:AFM:615257.615572}.

One challenging aspect is writing the code for the FMM operators, which depend on the choice of basis and kernel.
These operators generate the multipole expansions from particle distributions and manipulate those expansions to efficiently estimate the potential.
This complexity can grow significantly for hand optimized operators using even a moderate expansion order.
This may present itself as a stumbling block to developing and using FMM.

In this paper, we introduce the idea of using a symbolic algebra approach to automatically generate, and optimize, the FMM operator code for any kernel and any basis set up to any order.
This approach allows a developer to easily experiment with different forms and tailor the FMM operators to their specific needs.
The optimization process performs algebraic simplification and tedious subexpression elimination, which significantly reduces the total number of mathematical operations in the resulting code.
These unnecessary operations would otherwise be performed in a standard implementation of the operators.
Furthermore, we are able to symbolically verify the correctness of the expressions.
The implementation of this approach, written in Python, produces output for many common programming languages and is publicly available for download (see \ref{sec:Technical Details}).

We present our work in terms of the Cartesian basis set for the harmonic potential $1/r$.
This basis set is convenient for electrostatic applications that use point dipole and point quadrupole moments (see \ref{sec:Dipoles}), such as Polaris(MD) \citep{doi:10.1002/jcc.20932,doi:10.1002/jcc.21846,doi:10.1002/jcc.23237} and Tinker \citep{C7SC04531J}.
When constructing the multipole tensors in FMM, high order moments can be passed directly into the Cartesian representation without conversion.
As the Cartesian basis is already used by other FMM implementations such as ExaFMM \citep{doi:10.1260/1748-3018.7.3.301} our work could easily be incorporated into those applications as well.
The operators we generate are already used in the highly parallelized molecular dynamics software Polaris(MD).

Another consideration in choosing a basis set and FMM operators is the effect on computer memory usage and movement.
With modern supercomputer architectures, it is important to limit the movement of data internally within the memory hierarchy and externally over the network.
The number of stored coefficients impacts the performance as they must often be copied between machines. 
The memory required to store the multipole coefficients and the density of floating-point operations (FLOPs) per byte transferred are also important considerations.

The number of FLOPs required by the original spherical harmonics-based FMM algorithm \citep{1987JCoPh..73..325G} scales with the expansion order $p$ as $\O(p^4)$ and requires storing $\O(p^2)$ expansion coefficients ($2n+1$ terms per tensor rank, for $n \leq p$).
Later developments by the same authors using planewaves reduced that scaling to $\O(p^3)$ \citep{greengard_rokhlin_1997}. 
Using a Cartesian form of spherical harmonics combined with rotated translations, \cite{2014ComAC...1....1D} also demonstrated a method that scales as $\O(p^3)$.
\cite{doi:10.1260/1748-3018.7.3.301} suggests, however, that for low order expansions, the Cartesian basis has higher arithmetic complexity but smaller asymptotic constants, compared with spherical harmonics expansion with rotation, and may actually be more suitable for modern hardware.

A straightforward implementation in the Cartesian basis scales as $\O(p^6)$ and stores ${p+3}\choose{2}$ coefficients because of the introduction of many extraneous dependent terms in the multipole tensors.
However, using a traceless form of the tensor recovers the storage requirements of spherical harmonics.
At each expansion order the $n$-rank Cartesian multipole tensor $\ten M^{(n)}$ carries ${n+2}\choose{2}$ terms.
For our applications in parallel, distributed molecular dynamics simulations, where we choose to use the Cartesian basis set, this form reduces the storage and network communication by about 47\% for $p=7$ (see \tabref{tab:m elements} in the Appendix).

\cite{Applequist1989} discussed at length the traceless formalism and the intimate relationship between spherical harmonics and the traceless Cartesian multipole tensor.
There, a detracer operator was developed to convert any totally symmetric Cartesian tensor into a totally symmetric \emph{traceless} tensor.
Using this detracer, \cite{SHANKER2007732} derived equations for computing FMM interactions with traceless tensors, but the multipole shifting operator (which we will present later) does not actually yield another traceless multipole, making the method incomplete. 
More recent work by some of the same authors does not use the detracer (and therefore retains the entire non-traceless tensor), but instead takes advantage of the traceless nature of the harmonic gradient operator for $1/r$ to improve the local computational efficiency \citep{HUANG2018122}.
In \cite{Lorenzen:2012aa}, the authors present a complete traceless treatment of the FMM equations, but do not use the computationally efficient gradient operator as shown in \cite{Applequist1989}.

In \secref{sec:Code Gen}, we discuss the symbolic algebra generation and optimization procedure and our software tool \code{Mosaic} (Multipole Operators in Symbols, Automatically Improved and Condensed).
We have incorporated traceless tensors and the traceless gradient operator discussed in the literature into our code generator.
We also provide explicit expressions for all the FMM operators using the traceless formalism that translate easily into computer code.
In \secref{sec:Results} we compare three variants of the operators produced by our code.

To present a complete and consistent framework, we review the necessary mathematics in \ref{sec:Mathematical Background} and FMM approach in \ref{sec:Theory}.
Technical details of our particular FMM implementation are discussed in \ref{sec:Fast Multipole Method}.

\begin{figure}
\center\includegraphics[width=0.9\columnwidth]{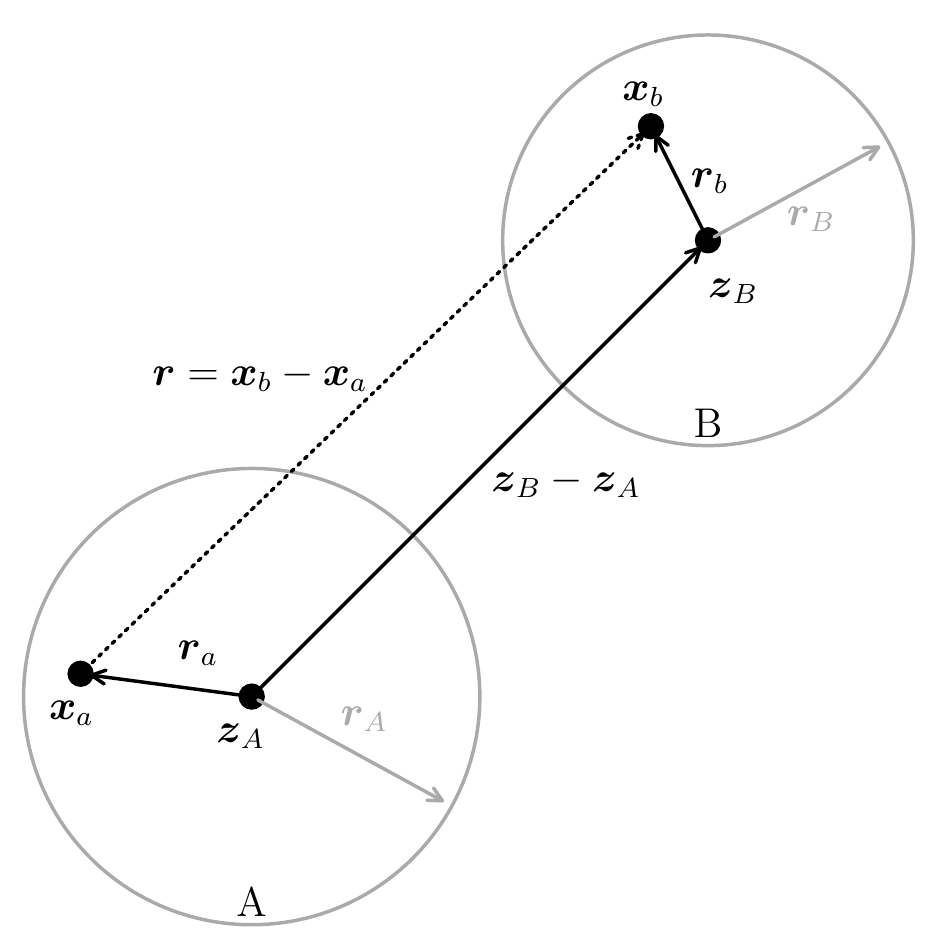}
    \caption{The vectors used by the multipole operators to compute the potential at $\vec x_b \in B$ due to a particle at $\vec x_a \in A$.  }
    \label{fig:setup}
\end{figure}

\section{Code Generator\label{sec:Code Gen}}

\subsection{FMM Operator Overview}

To motivate our discussion of the symbolic algebra approach, we present the FMM operators we currently use as input to the code generator.
The generator is flexible enough to accept other forms for a different choice of basis and kernel.
We use the Cartesian basis set as discussed in the introduction and the traceless form of the multipole tensors.
In the Appendices we review the mathematical background of these operators, which is based on the work of \cite{Applequist1989}, \cite{2002JCoPh.179...27D}, and \cite{SHANKER2007732}.

We use the Green's function $\phi(\vec r = \vec x_b - \vec x_a) = |\vec r|^{-1}$ found in electrostatics or gravity calculations.
Because it is harmonic, we can take advantage of the traceless gradient operator as shown in \cite{HUANG2018122}.
A diagram of vectors is presented in \figref{fig:setup}.
In the following expressions, tensors with an overbar are traceless and the multi-index notation is used to reference scalar components of the tensors (see \ref{sec:mi notation}).
The resulting equations are all sums of products of scalar values.

\smallskip\noindent\emph{Operator P2M (Particle to multipole). Create a multipole expansion from a collection of particles.}
The traceless multipole tensor for particles in an enclosed region centered at $\vec z_A$ is
\begin{align} \label{eq:P2M}
    &\tl{\ten M}_{\mi n}(\vec z_A = \vec x_a - \vec r_a) = \sum_{a} q_a \D[\ten M']_{\mi n} \\
    &\ten M'_{\mi n} = -\vec r_a^{\mi n}
\end{align}
where each particle $a$ has charge $q_a$ and position $\vec x_a$ in the region, and $\D$ is a detracer operator (see \ref{sec:Mathematical Background}).

\smallskip\noindent\emph{Operator M2M (Multipole to multipole). Recenter a multipole expansion.}
The M2M operator shifts the expansion center of an existing traceless multipole expansion at $\vec z$ to a new location $\vec z'$:
\begin{align} 
\label{eq:M2M}
    &\tl{\ten M}_{\mi n}(\vec z' =  \vec z - \vec \delta) = \D[{\ten M'}]_{\mi n} \\
\label{eq:M2M part 2}
    &{\ten M'}_{\mi n}(\vec z') = \sum_{|\mi k| \leq n} 
    \frac{\mi n!}{\mi k!(\mi n-\mi k)!}  
    (-\vec \delta)^{\mi n - \mi k} \tl{\ten M}_{\mi k}(\vec z)
\end{align}
The factor of $\mi n!/ \mi k!$ accounts for symmetries in the tensor product of $\ten{\delta}$ with $\tl{\ten M}$ when using multi-index notation%
\footnote{
It is tempting to remove this by absorbing $\mi k!$ into the definition of $\tl{\ten M}$.
However, if $\tl{\ten M}$ is traceless this will break the harmonic property (consider for $\mi k = (0,0,3)$ that $\ten M_{zzz}/3! = -(\ten M_{xxz} + \ten M_{yyz})/3! \ne -(\ten M_{xxz}/2! + \ten M_{yyz}/2!)$).
One could work around this, but the resulting calculation outweighs the extra multiplications.
}%
.
The shift itself does not result in a traceless tensor until we apply the detracer $\D$.

We find that the overall number of FLOPs is reduce by inserting \eqnref{eq:M2M part 2} into \eqnref{eq:M2M}, expanding the detracer, and rearranging terms to arrive at
\begin{align} \label{eq:Detracer2}
    &\tl{\ten M}_{\mi n}(\vec z' =  \vec z - \vec \delta) =
            \tl{\ten M}_{\mi 0}(\vec z)
            \T[-\vec \delta^n]_{\mi n} + 
      \frac{1}{(2n-1)!!} \times  \nonumber \\
    &
        \left[
        \sum_{1 \leq |\mi q| \leq n} \frac{\tl{\ten M}_{\mi q}(\vec z)}{\mi q!}
    \sum_{\mi m \leq \lfloor \mi n/2 \rfloor} 
        f_{\mi n, \mi m} \;
        \sum_{|\mi k|=m} 
        \frac{m!}{\mi k!}\frac{(\mi j + \mi q)!}{\mi j!}
        \T[-\vec \delta^j]_{\mi j} \right]
\end{align}
where $\mi j = \mi n + 2\mi k - 2\mi m - \mi q$ and the traceless tensor $\T[-\vec \delta^i]$ has been used.

\smallskip\noindent\emph{Operator M2L (Multipole to local). Compute a local field tensor located outside a multipole expansion.}
The local field tensor 
\begin{equation} \label{eq:M2L}
    \tl{\ten L}_{\mi n}(\vec z' = \vec z + \vec r) = \sum_{|\mi m| \leq p-n} \frac{1}{\mi m!}\tl{\ten M}_{\mi m}(\vec z) \tl{\ten D}_{\mi n + \mi m}(\vec r)
\end{equation}
describes the potential field at $\vec z'$ produced by a multipole expansion $\tl{\ten M}(\vec z)$.
It is only valid outside the region defining $\tl{\ten M}$.
The traceless gradient $\tl{\ten D}_{\mi k}(\vec r) = \vec\nabla^{\mi k}r^{-1}$  is computed efficiently using \eqnref{eq:traceless grad}.

\smallskip\noindent\emph{Operator L2L (Local to local). Relocate a local field tensor.}
The center of a local field tensor $\tl{\ten L}$ can be changed using
\begin{equation} \label{eq:L2L}
    \tl{\ten L_{\mi n}}(\vec z' \approx  \vec z + \vec \delta) = \sum_{|\mi k| \leq p-n} \frac{\vec \delta^{\mi k}}{\mi k!} \tl{\ten L}_{\mi n + \mi k}(\vec z)
\end{equation}
Unlike \op{M2M}, this operation is \emph{not exact} because the summation is bounded by the expansion order.
An exact translation would require the sum to go to infinity.

\smallskip\noindent\emph{Operator L2P: (Local to particle). Relocate a local field tensor at a particle position.}
Typically used for computing only (derivatives of) the potential at a particle position $\vec x_b$. 
We only need to shift the field tensor to the location of the particle and examine the appropriate tensor components:
\begin{equation} \label{eq:L2P}
		\nabla^{\mi n}
    \Phi(\vec x_b = \vec z_B + \vec r_b) 
    \approx \tl{\ten L}_{\mi n}(\vec x_b)
\end{equation}
From our definitions, the potential is defined to be positive. For astrophysical applications, the potential and respective gradients should be negated.


\subsection{Generating Computer Code from Symbolic Algebra}

\begin{figure*}
\includegraphics[width=\textwidth]{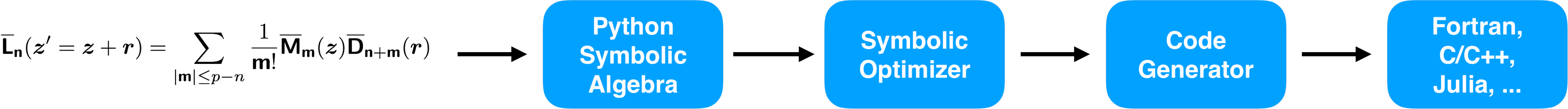}%
    \caption{
        The code generator \code{Mosaic} is a pipelined process that converts a expressions for a fast multipole method operator, such as \op{M2L} pictured here \eqnrefp{eq:M2L}, into optimized code in any number of programming languages. The mathematical expression is written in Python using the symbolic representation of the Python library SymPy. 
This is then manipulated through algebraic simplification and subexpression elimination to reduce the total number of floating-point operations in each operator.
The final form is then converted to routines for the desired programming language.
Currently, Fortran and \LaTeX are possible but common programming languages supported by Sympy (e.g., C/C++, Fortran, Julia, etc.) can be easily be included.
    }
\label{fig:flow chart}
\end{figure*}

In this section we present our novel idea to first construct the fast multipole method operators in symbolic algebra and then convert this representation into optimized computer code (\figref{fig:flow chart}).
We call the tool implementing this approach \code{Mosaic} (Multipole Operators in Symbols, Automatically Improved and Condensed).

This allows great flexibility for the developer to choose and change the FMM basis set, kernel, and expansion order.
Different choices are easy to compare and the developer can tailor the implementation to their specific needs.

The symbolic algebra approach also simplifies the operator programming and verification process.
For the FMM operators presented above, the symbolic representation in the generator is designed to closely match these expressions.
The generator outputs expressions for each tensor component and optimizes these through algebraic manipulation and common subexpression extraction.
Subexpression extraction is applied across the entire set of tensor components.
The pre-optimized output of the generator allows the developer to verify correctness for low expansion orders and then generate production code for higher orders.

Self-consistent validation can also be performed automatically.
For instance, shifting a multipole with M2M is expected to produce the same result as a multipole created using P2M at the new, shifted location.
The generator performs this manipulation symbolically and can compare the two resulting expressions.
Through numerical substitution, we can also verify correctness by comparing the numerical output to known quantities at the time the operators are generated and before the code is used in a simulation.

Our implementation is currently designed for the Cartesian basis set and can use fully traceless and non-traceless forms as well incorporate the traceless gradient optimizations presented in \cite{HUANG2018122}.
\code{Mosaic} is implemented in Python using the SymPy symbolic Python library \citep{10.7717/peerj-cs.103} and outputs results in many common computer languages---such as C/C++, Fortran, Julia, etc.---with minor modifications.

Without the tool we have developed, the programmer must choose either to implement the operators as nested loops or hand written expressions for each tensor component.
The loop approach is compact, and easy for a compiler to vectorize, but difficult to reuse common elements between components.
The hand written approach is tedious and error-prone to write, but allows for efficient optimizations, even though it quickly becomes overwhelming at higher order.
However, the expansion order, choice of basis, and kernel are fixed.
The astrophysics code \textsc{gyrfalcON} \citep{2002JCoPh.179...27D} uses recursive C++ template programming to generate the FMM operator expressions at compile time.
This is simpler than the previously mentioned methods, but is limited to C++, is difficult to heavily optimize, and for large expansion orders is time consuming to compile.

Our approach allows for flexible experimentation or testing or to produce a known correct form as a starting point for further development.
If another form of output is desired (e.g., coded loops) a first pass with the generator can ensure the correctness of the approach.

The computer code generator in \code{Mosaic} is a pipelined process as shown in \figref{fig:flow chart}.
In the first stage, the equations for each operator are recreated in Python in a form that mirrors the equations above.
For example, the code for \op{M2L} \eqnrefp{eq:M2L} in Python is
\begin{verbatim}
L = OrderedDict()
for n in indices(0,p):
    L[n] = 0
    for m in indices(0,p-sum(n)):
        L[n] += Msym[m] * Dsym[miadd(n,m)]
\end{verbatim}
The variables $\tt n$ and $\tt m$ are tuples in the multi-index format and $\tt Msym$ and $\tt Dsym$ are dictionaries of symbols for the tensor components.
The $\tt indices$ function returns a list of tuples whose magnitudes span the requested range.
The typical output in symbols for the element corresponding to $\ten L_x$ looks like
\begin{verbatim}
L[(1,0,0)] = D_x*M_0 + D_xx*M_x + D_xxx*M_xx/2 
  + D_xxy*M_xy   + D_xxz*M_xz + D_xy*M_y 
  + D_xyy*M_yy/2 + D_xyz*M_yz + D_xz*M_z 
  + D_xzz*M_zz/2
\end{verbatim}

While no explicit optimizations take place here, the symbolic library automatically performs some algebraic manipulation such as eliminating terms with factors of zero.

In the second stage, we perform the optimizations.
Because this work is performed ahead of compilation, complex, time consuming, manipulations are possible.
Constants are factored out and the common subexpression elimination feature of SymPy is applied.
All equations for a single operator are examined together to find repeated subexpressions.
These are extracted into new variables that are reused many times.
The procedure is applied iteratively until no new changes are possible.
Typically this is achieved by four iterations.
Any expression with an exponent is expanded into multiplications to allow the compiler to prefer multiplication over calling a power function.
Finally, any constant rational expressions are explicitly replaced with their floating-point equivalent. 

In the third stage, the final symbolic form of all the operators is written in the chosen computer language along with any necessary header files.

As an example of the optimization step, consider the following excerpt from \op{M2L} at $p=5$.
Before optimization the diagonal components of the gradient operator {\tt D\_xx, D\_yy, D\_zz} are computer with
\begin{verbatim}
h       = x**2 + y**2 + z**2
u2      = 1.0/h
u       = sqrt(u2)
D_xx    = -u**5*h + 3*u**5*x**2
D_yy    = -u**5*h + 3*u**5*y**2
D_zz    = -(D_xx + D_yy)
\end{verbatim}
where $\tt h$ is the squared distance over which \op{M2L} is applied.
Since the gradient is traceless, the {\tt D\_zz} term is computed in terms of the components along the diagonal.
The optimized version converts this into
\begin{verbatim}
c1      = z
b1      = y
a1      = x
a2      = a1*a1
v0      = a2
v5      = 3.0d0*v0
b2      = b1*b1
v1      = b2
v7      = 3.0d0*v1
c2      = c1*c1
h       = c2 + v0 + v1
v4      = -h
u2      = 1.0/h
u       = sqrt(u2)
u3      = u*u2
v2      = u3
u5      = u2*u3
v3      = u5
D_xx    = v3*(v4 + v5)
D_yy    = v3*(v4 + v7)
D_zz    = -(D_xx + D_yy)
\end{verbatim}

All explicit powers have been eliminated and several variables are reused.
While the difference in FLOPs is small for this example (23 vs. 15), over the entire function the number of FLOPS is reduce from 1228 to 986.

There are no loops in the final code.
This can make it difficult for an optimizing compiler to take advantage of vector instructions.
However, in FMM one can often collect multiple interactions and apply the same operator at once \citep{2014ComAC...1....1D}.
With \code{Mosaic}, it is a simple matter to generate specialized routines that accept multiple packed tensors and use vectorized code to process them.
Our tests have shown a 1.8x speedup for \op{M2L} at $p=7$ using two packed tensors (the speed-up is less than 2x due to the packing/unpacking of the tensors).
For newer processors that support vector instructions with four or more floating-point numbers, new routines are simple to generate. 

\begin{table*}[t]
    {\footnotesize
    \begin{tabular*}{\textwidth}{l|@{\extracolsep{\fill}}rr|rr|rr|rrr}
Operator & \multicolumn{6}{c|}{Number of FLOPs} & \multicolumn{3}{c}{Timing [ns]} \\
\hline
$p=3$ & \multicolumn{2}{c}{FT} & \multicolumn{2}{c}{TG} & \multicolumn{2}{c|}{AP} & FT & TG & AP \\
\hline
P2M                  & 34 & (61) & 41 & (54) & 18 & (19) & 9.54 & 9.57 & 4.88  \\
M2M                  & 205 & (309) & 137 & (184) & 25 & (25) & 35.97 & 22.97 & 5.35  \\
M2L                  & 213 & (258) & 201 & (249) & 118 & (166) & 43.15 & 42.26 & 30.57  \\
L2L                  & 122 & (172) & 122 & (172) & 78 & (109) & 19.65 & 19.78 & 11.69  \\
L2P                  & 123 & (173) & 123 & (173) & 64 & (79) & 19.40 & 19.49 & 9.32  \\
\hline
$p=5$ & \multicolumn{2}{c}{FT} & \multicolumn{2}{c}{TG} & \multicolumn{2}{c|}{AP} & FT & TG & AP \\
\hline
P2M                  & 124 & (293) & 109 & (213) & 69 & (143) & 24.57 & 26.00 & 14.74  \\
M2M                  & 1300 & (2803) & 733 & (1399) & 403 & (823) & 280.12 & 136.74 & 69.89  \\
M2L                  & 986 & (1228) & 905 & (1173) & 700 & (957) & 227.80 & 197.06 & 156.08  \\
L2L                  & 520 & (1174) & 520 & (1174) & 456 & (948) & 97.36 & 97.96 & 82.60  \\
L2P                  & 426 & (946) & 426 & (946) & 258 & (404) & 76.42 & 76.23 & 46.52  \\
\hline
$p=7$ & \multicolumn{2}{c}{FT} & \multicolumn{2}{c}{TG} & \multicolumn{2}{c|}{AP} & FT & TG & AP \\
\hline
P2M                  & 342 & (930) & 229 & (532) & 222 & (539) & 58.79 & 55.80 & 38.29  \\
M2M                  & 5997 & (13934) & 2706 & (6216) & 2372 & (5779) & 1564.61 & 622.58 & 560.80  \\
M2L                  & 3158 & (3923) & 2881 & (3737) & 2553 & (3362) & 910.01 & 794.47 & 707.95  \\
L2L                  & 1404 & (4652) & 1404 & (4652) & 1300 & (4100) & 311.05 & 312.87 & 285.24  \\
L2P                  & 887 & (2837) & 887 & (2837) & 588 & (1154) & 193.21 & 190.60 & 126.39  \\
\hline
\end{tabular*}
    }
    \caption{Representative results from the code generator for the five FMM operators at three expansion orders $p=(3,5,7)$. 
As discussed in \secref{sec:Results}, three variants of the operators were generated---FT (fully traceless), TG (traceless gradient), and AP (astrophysics).
The number of floating-point operations (FLOPs) for the optimized code is given in the first column and in parenthesis the number of FLOPs in the unoptimized version (-NoOpt).
The last three columns present timing results for the optimized case.
For the FT and TG variants, the operators \op{L2L} and \op{L2P} are expected to be identical.}
    \label{tab:fmm op table}
\end{table*}

\section{Discussion and Analysis\label{sec:Results}}
Using the code generator \code{Mosaic}, we produced three different variants of the FMM operators in both optimized and non-optimized form.
The traceless gradient (TG) variant implements the optimization suggested in \cite{HUANG2018122} to use the traceless gradient operator.
The multipole tensor is not traceless and all components must be stored.
However, the \op{M2L} produces a traceless local field tensor and subsequent operators \op{L2L} and \op{L2P} are traceless.
The fully traceless (FT) form uses the traceless multipole tensor and the traceless gradient operator.
We therefore only need to store the independent terms of the multipole tensor and recompute dependent terms if necessary.
Both the FT and TG variants calculate the monopole and quadrupole (hessian) terms in \op{L2P}.
The monopole term, which corresponds to the scalar potential, adds a substantial number of FLOPs to each operator and can be removed if necessary.
The third version is intended for astrophysical (AP) applications typically found in cosmological or galaxy simulations.
In gravity codes, the expansion is normally located at the center of mass, which we assume here. 
The dipole vanishes in this case and the number of operations in the higher rank tensors is reduced. 
Furthermore, only the gradient of the potential $\nabla\Phi$ is required, removing the need for the monopole and quadrupole to be calculated in \op{L2P}.
We use the fully traceless approach in AP to follow the work of \cite{2001PhDT........21S} used in \textsc{Pkdgrav}.

Each of these three variants also has an unoptimized version (-NoOpt) where the majority of complex optimizations have not been performed.
Only simple factorization of constants is done.

In \tabref{tab:fmm op table} we list the number of floating-point operations (additions, multiplications, etc.) produced for each operator in each variant at expansion orders $p={3,5,7}$.
The most critical operator is \op{M2L}, which is used far more often than the other operators (up to 60x in our experiments) .
The range of expansion order covers the common use cases found in production codes that do not require extremely high accuracy.
For many electrostatic simulations where cells may be charge neutral, and the expansion dominated by the dipole term, an expansion order up to $p=7$ is necessary to achieve an accurate calculation.
Astrophysical applications do not suffer from this issue as the mass must always be positive and the multipole expansion is already reasonably accurate at low order, although some implementations of FMM for astrophysics use $p=5$ \citep[e.g.][]{2017ComAC...4....2P}.

The numbers presented here are for reference only and should not be interpreted as the minimal number of operations possible.
Hand optimization may lead to a further reduction and future improvements to the code generator and SymPy may also be possible.
Indeed, \cite{2017ComAC...4....2P} claims 473 operations for \op{M2L} at $p=5$ in Table 3, while we generate 700.
However, these operators in that work are only implemented for a single expansion order.
Compared to the operators in \cite{doi:10.1260/1748-3018.7.3.301}, which have been highly optimized for low $p$, we are less than a factor of 2 slower, but it is unclear if this author has computed the monopole term, discussed above.

\begin{figure}
\includegraphics[width=\columnwidth]{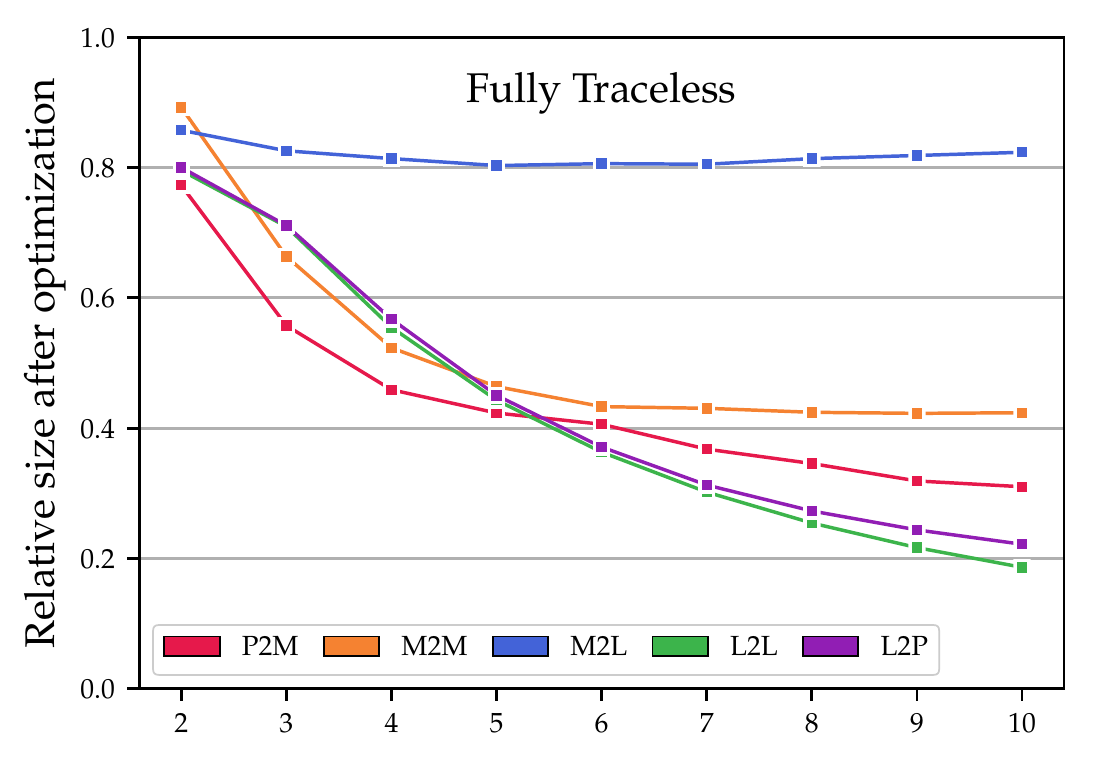}%
    \caption{
    The relative size of each FMM operator (measured in FLOPs) after being processed by the symbolic algebra optimizer in the fully traceless (FT) case.
The optimizer performs algebraic manipulation on the symbolic representation of each FMM operator and extracts common subexpressions into new variables that are computed once and reused. 
Since \op{M2L} uses the traceless gradient, the operator is nearly in an ideal form.
Nonetheless, the optimizer is able to reduce the total number of operations by 20\%.
    }
\label{fig:compare optimizations}
\end{figure}

In \figref{fig:compare optimizations} we plot the relative reduction in FLOPs after running the symbolic code optimizer in the FT case.
The other variants produce similar results.
Recall that the optimizer extracts common subexpressions into temporary variables and performs other mathematical reductions.
For \op{M2L}, the non-optimized version is already in a reduced form through the use of the traceless gradient operator and therefore we only observe a 20\% improvement.
For the other operators, however, we see a 40-60\% reduction for $p \ge 5$.
It should also be noted that the non-optimized number of operations is increasing substantially with increasing order, yet we observe a nearly constant level of improvement at higher order, suggesting that the optimizer is finding more redundancy at higher order.

\begin{figure}
\includegraphics[width=\columnwidth]{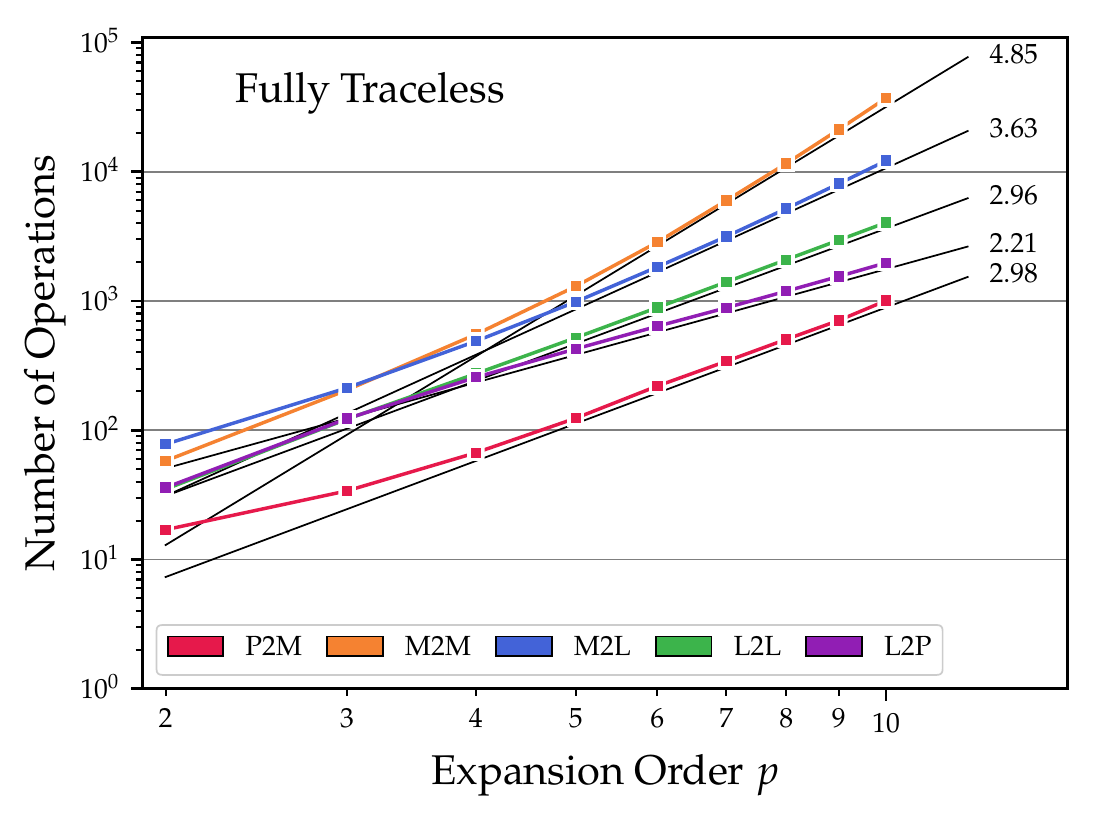}
\includegraphics[width=\columnwidth]{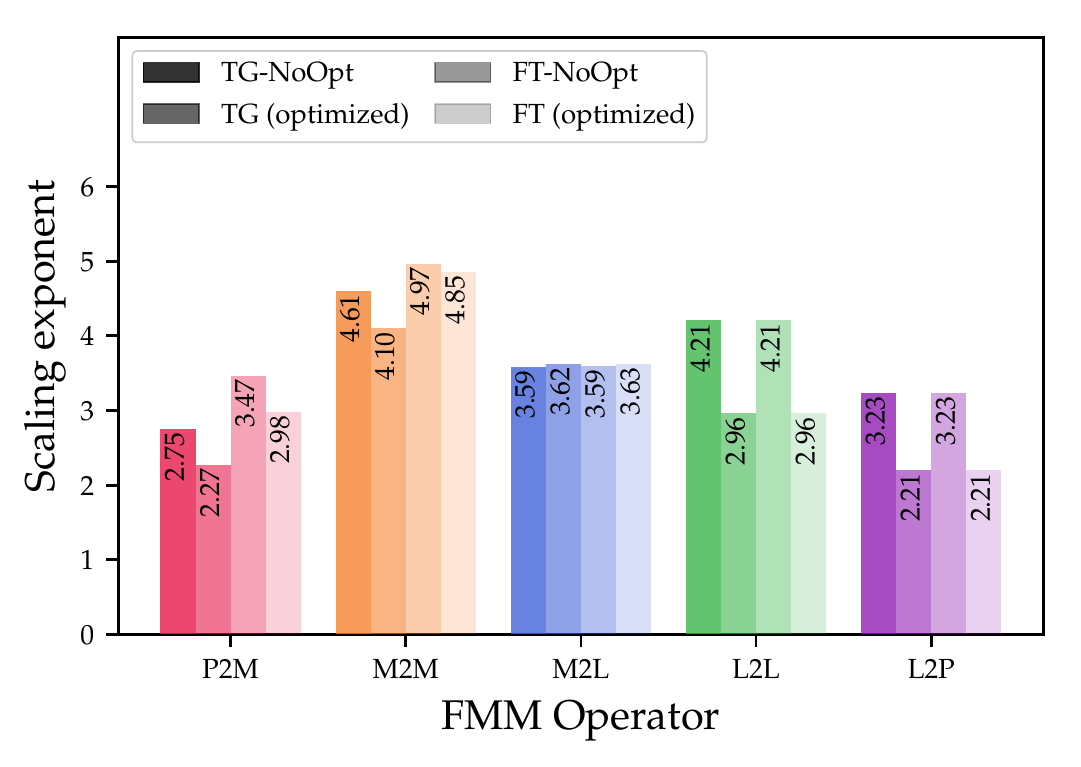}%
    \caption{
\emph{Top:}~The number of floating-point operations generated for each optimized FMM routine in the full traceless (FT) variant as a function of the expansion order $p$. 
Trend lines have been fit using the power-law $p^m$ over the range $5 \le p \le 10$ to determine an empirical scaling.
Each trend line is labeled with the corresponding $m$.
The code optimizer not only reduces the total number of mathematical instructions in each operator compared with the expected number from the equations, it
can also improve the empirically measured scaling properties.
\emph{Bottom:} We compare the scaling exponents derived for each of the FMM operators as above.
The fitted exponent $m$ is given in each bar.
We show four variants of an operator from left to right: 1) TG-NoOpt, the traceless gradient, non-optimized version, 2) TG, the traceless gradient after symbolic optimization, 3) FT-No-Opt, the fully traceless, non-optimized version, and 4) FT, the traceless optimized version.
    }
\label{fig:op scaling}
\end{figure}

In the top plot of \figref{fig:op scaling} we show the number of operations after optimization as a function of expansion order for the fully traceless (FT) case.
Each curve is fit with a simple power-law $p^m$ over the range $5 \le p \le 10$ to determine an empirical scaling.
The power-law exponent $m$ is written next to each trend line.
In the bottom plot of \figref{fig:op scaling} we compare the fitted exponents for both FT and TG in the optimized and unoptimized forms.
The scaling is observed to improve after optimization for all operators except \op{M2L}, where the scaling remains nearly constant.
The result is surprising, since one may have expected optimization would reduce the overall number of operations but not change the scaling exponent.
We suspect that the optimizer is identifying (although not knowingly) further symmetries in the equations that can be exploited.
However, despite investigating this aspect further, we are unable to confirm this and will defer this to future work. 

%


\section{Conclusions\label{sec:Conclusions}}
We have developed an optimizing code generator called \code{Mosaic} for FMM operators based on a symbolic algebra library.
The symbolic approach can easily produce code for any expansion order and can be extended to use any basis or kernel function.
The method applies algebraic modifications to reduce the number of floating-point operations and can symbolically verify correctness.

We demonstrated the method using three variations all based on the Cartesian basis for the $1/r$ potential kernel, commonly found in electrostatics and gravity simulations. 
The traceless gradient (TG) variant stores the complete multipole tensor but uses the traceless property of the harmonic potential.
The fully traceless (FT) variant additionally uses the traceless multipole tensor, which reduces memory usage and communication, but incurs an overhead of needing additional operations to create the traceless multipole tensor.
Our final variant is optimized for astrophysics (AP) applications that can eliminate computation by assuming the dipole term vanishes.

We find that the optimizer is able to reduce the number of operations compared with a direct implementation of the equations by 20-80\%, depending on the operator.
We have further observed that the optimized code improves the scaling for many FMM operators over the range of expansion orders tested.

Producing the code for the fast multipole operators is one of the major stumbling blocks to developing FMM.
Since the symbolic algebra approach automates this aspect, the developer is free to experiment with different basis sets or kernels with ease.
The software tool is now available for public download.


\section{Acknowledgements}
JPC would like the thank Walter Dehnen for several lively and educational discussions.
The authors also thank Martin Zacharias and Michel Masella for their comments on this manuscript.
The referee comments have been invaluable to improving the focus and quality of this article.
Performance tests were conducted on the in-house cluster of the Lehrstuhl f\"ur Theoretische Biophysik at TUM.

\appendix


\section{Mathematical Background\label{sec:Mathematical Background}}
Much of the mathematical basis of our work is derived from \cite{Applequist1989}, which contains full proofs and additional details.
Here we summarize the main results used in this article.
We simplify the notation using the multi-index tuple reviewed in \ref{sec:mi notation}.
In the three dimensional Cartesian basis, an $n$-rank tensor $\ten A^{(n)}$ contains $3^n$ elements. 
We assume the tensor is symmetric so that only the upper triangle of the tensor is necessary.
The remaining $(n+1)(n+2)/2$ elements are stored in a \emph{compressed form} where an element is $A^{(n)}(n_1,n_2,n_3)$.
The polytensor $\ten A = \{\ten A^{(0)}, \ten A^{(1)}, \ten A^{(2)}, \dots\}$ is a set of tensors of increasing rank that will be used to store the tensors of multipole moment expansion used by the fast multipole method.

The multi-index notation naturally captures the compressed format so $\ten A^{(n)}_{\mi n} \equiv A^{(n)}(n_1,n_2,n_3)$.
If the superscript is omitted then $\ten A_{\mi n}$ is the scalar element at index $\mi n$ in the tensor $\ten A^{(n)}$ in the polytensor $\ten A$.

The $n$-fold contraction of two tensors $\ten C^{(m)} = \ten A^{(m+n)} \cdot n \cdot \ten B^{(n)}$ is computed as
\begin{equation}
    \ten C^{(m)}_{\mi m} = \sum_{|\mi k|=n} \frac{n!}{\mi k!}\ten A^{(m+n)}_{\mi m+\mi k} \ten B^{(n)}_{\mi k}
\end{equation}
where the summation is over all possible values of $\mi k$ such that $|\mi k|=n$.
While the contraction $\ten A^{(m+n)} \cdot n \cdot \ten B^{(n)}$ produces a tensor of reduced rank $m$, the direct product (also known as a Kronecker product) $\ten C^{(m+n)} = \ten A^{(m)}\ten B^{(n)}$ is a $(m+n)$-rank tensor.
The direct product of $n$ copies of a vector is a symmetric $n$-rank tensor. 
For a vector $\vec r$, $\vec r^k_{\mi n}$ is the ${\mi n}$th element of the tensor $\vec r^k$.

The direct product between two symmetric tensors is not, in general, symmetric and cannot be represented using the multi-index notation
(consider for example the simple direct product of $[x,y] [a,b] = [[xa,xb], [ya,yb]]$).
For symmetric tensors $\ten A^{(m)}, \ten B^{(n-m)}$ the symmetric direct product $\mathrm{Sym}(\ten C^{(n)}) = \ten A^{(m)} \ten B^{(n-m)}$ using the multi-index notion is
\begin{equation}\label{eq:Sym}
    \mathrm{Sym}(\ten C^{(n)}_{\mi n})
        = \sum_{|\mi k| = |\mi n|}
        \frac{k!}{\mi k!} \frac{(n - k)!}{(\mi n - \mi k)!} \frac{\mi n!}{n!}
        \ten A^{(m)}_{\mi k} 
        \ten B^{(m-n)}_{\mi n - \mi k}
\end{equation}

The trace in one index pair of a tensor $\ten A^{(n)}$ is a reduced tensor of rank $k=n-2$ and is defined to be $\ten A_{\mi k}^{(n:1)} = \ten A_{\mi k + (2,0,0)} + \ten A_{\mi k + (0,2,0)} + \ten A_{\mi k + (0,0,2)}$.
A tensor is totally traceless if the trace vanishes regardless of $\mi k$.
A totally symmetric tensor which is traceless for some $\mi k$ is traceless for all $\mi k$ and is said to be totally symmetric and traceless. 
Similarly, the $m$-fold trace is $\ten A^{(n:m)}_{\mi k} = \sum_{|\mi j|=m} \frac{m!}{\mi j!}\ten A_{\mi k + 2\mi j}$, where $k=n-2m$.

A totally symmetric tensor $\ten A^{(n)}$ is converted to a totally symmetric traceless tensor $\T \ten A^{(n)}$ using the detracer operator $\T$.
The elements of $\T \ten A^{(n)}$ are 
\begin{align} \label{eq:Detracer}
\noindent
    \T[\ten A^{(n)}]_{\mi n}  
    &= \sum_{\mi m \leq \lfloor \mi n/2 \rfloor} f_{\mi n, \mi m} \; \ten A^{(n:m)}_{\mi n-2\mi m} \nonumber \\
    &= \sum_{\mi m \leq \lfloor \mi n/2 \rfloor} \left[ f_{\mi n, \mi m} \sum_{|\mi k| = m} \frac{m!}{\mi k!} \ten A_{\mi n-2\mi m + 2\mi k} \right]
\end{align}
where
\begin{equation}
    f_{\mi n, \mi m}  = (-1)^m (2n-2m-1)!! \frac{\mi n!}{2^m \mi m! (\mi n-2\mi m)!} \\
\end{equation}
If $\ten B^{(n)} = \T \ten A^{(n)}$ then
$\T \ten B^{(n)} = (2n-1)!! \ten B^{(n)}$, 
which is not idempotent ($\T\T \neq \T$). 
We define another operator $\D\ten A^{(n)} \equiv \T\ten A^{(n)} / (2n-1)!!$ that satisfies $\D\D=\D$.
(This naming convention follows that of \cite{Applequist1989}, which differs from \cite{SHANKER2007732}).

For $\ten A^{(n)}$ and $\ten B^{(n)}$ totally symmetric there are several useful identities.
In particular, the detracer exchange theorem:
\begin{equation}\label{eq:T exchange}
    \ten A^{(n)} \cdot n \cdot \T \ten B^{(n)} = \T \ten A^{(n)} \cdot n \cdot \ten B^{(n)}
\end{equation}
and
\begin{equation}
    \T \ten A^{(n)} \cdot n \cdot \T \ten B^{(n)} = (2n-1)!! \ten A^{(n)} \cdot n \cdot \T \ten B^{(n)}
\end{equation}
\cite{SHANKER2007732} demonstrated that a contraction where at least one tensor is traceless is also traceless.
However, the direct product of a non-traceless tensor with a traceless tensor is not, in general, traceless.

The gradient $\vec\nabla^{\mi n} r^{-1}$ is intimately related to $\T \vec r^n$.
As early as Maxwell, but then later in \cite{APPLEQUIST1984279,Applequist1989} and \cite{doi:10.1080/00268978100102251}, it was shown that
\begin{equation} \label{eq:traceless grad}
    \vec\nabla^{\mi n}r^{-1} = (-1)^n r^{-2n-1} \T[\vec r^n]_{\mi n} 
\end{equation}
Using the identity
\begin{equation} 
    r^{2m} = \sum_{|\mi k|=m} \vec r^{2\mi k}\frac{m!}{\mi k!}
\end{equation}
the traceless tensor $\T[\vec r^n]$ assumes a compact and computationally efficient form:
\begin{equation} \label{eq:Detracer3}
    \T[\vec r^n]_{\mi n} = \sum_{\mi m \leq \lfloor \mi n/2 \rfloor} 
    f_{\mi n, \mi m} \;
    r^{2m} \vec r^{\mi n-2\mi m}  
\end{equation}
that contains only a single summation instead of two for the general case \eqnrefp{eq:Detracer}.


\section{Physical Theory\label{sec:Theory}}
The $n$th-order multipole moment of a distribution $\rho$ centered at the origin is defined as
\begin{equation} \label{eq:multipole moment}
    \ten \mu^{(n)} = \frac1{n!}\int_v \rho(\vec s) \vec s^n \mathrm{d} v
\end{equation}
where the integration is over points $\vec s$ in a finite volume sphere $v$. 
For any point $\vec r$ outside the sphere the potential arising from $\rho(\vec s)$ with
Green's function $\phi$ is given by the multipole expansion
\begin{equation} \label{eq:multipole expansion}
    \Phi(\vec r) = \sum_{n=0}^{\infty} (-1)^n \ten \mu^{(n)} \cdot n \cdot \nabla^n \phi
\end{equation}
The potential energy of charge $q$ at $\vec r$ is $U(\vec r) = q\Phi(\vec r)$.
Consider a distribution of discrete particles with individual weights $q_i$ (e.g., charge or mass).
Then in distinct, non-overlapping regions $A,B$, with centers $\vec z_A,\vec z_B$ respectively, the total potential at $\vec x_b \in B$ due to $A$ is
\begin{equation}\label{eq:PhiAB}
\Phi_{A\rightarrow B} = \sum_{a \in A} q_a \phi(\vec x_b - \vec x_a)
\end{equation}
In \figref{fig:setup} we sketch the general setup of two interacting regions and the vectors.
Define $\vec r_a = \vec x_a - \vec z_A$, $\vec r_b = \vec x_b - \vec z_B$, and $\vec r = \vec z_B - \vec z_A$, so that $\vec x_b - \vec x_a = \vec r + \vec r_b - \vec r_a$.
The Taylor expansion of $\phi$ to order $p$ in both $\vec r_a$ and $\vec r_b$ is given by the expression
\begin{equation}\label{eq:phi expansion}
    \phi
     \approx
     \sum_{|\mi n| \leq p}\sum_{|\mi m| \leq p-n} \frac{ (-1)^{|\mi m|}\vec r_a^{\mi m}\vec r_b^{\mi n} }{\mi n!\mi m!}\vec\nabla^{\mi n + \mi m} \phi(\vec r)
\end{equation}

We pull apart \eqnref{eq:PhiAB} to define the operators used by the fast multipole method (described in \ref{sec:Fast Multipole Method}).
These operators are often presented in terms of tensor contractions in the literature but here we use the notation from \ref{sec:Mathematical Background} to make the presentation clearer and closer to a programming implementation.

The following discussion now assumes the Green's function $\phi = \vec r^{-1}$ so that we can exploit the traceless property of the gradient operator when the using this form.
The operators use the fully traceless formalism found in \cite{SHANKER2007732} (although we present a correction to the operator \op{M2M} that maintains the traceless property of the multipole expansion).
Tensors labeled with a bar such as $\tl{\ten A}$ are totally symmetric and traceless. 
It should be understood that only the independent elements of a traceless tensor are stored. 
Where \emph{dependent} elements are needed, they are calculated on demand (and reused) from the independent elements of the trace (e.g., $\ten A_{zz} = -\ten A_{xx} - \ten A_{yy}$).

The traceless multipole expansion for a discrete set of particles in a region $A$ centered at $\vec z_A$ is given by 
 
\begin{align} \label{eq:multipole v1}
    &\tl{\ten M}_{\mi n}(\vec z_A = \vec x_a - \vec r_a) = \D[\ten M']_{\mi n} \\
    &\ten M'_{\mi n} = \sum_{a \in A} q_a \frac{(-1)^{n}}{n!}\vec r_a^{\mi n} 
\end{align}
for all $|\mi n|$ no greater than the expansion order $p$.
This is reminiscent of the continuous form in \eqnref{eq:multipole moment} with change of sign from \eqnref{eq:multipole expansion} absorbed into the definition.
Note that the multipole expansion object $\tl{\ten M}$ is a polytensor that contains tensors of all ranks up to $p$.

The field generated by $A$ at a point $\vec z'$ outside $A$ is described by the local field tensor $\tl{\ten L}(\vec z')$.
It is the contraction between $\tl{\ten M}$ and $\vec\nabla^{\mi k}\phi$ and is only valid outside the volume defining $\tl{\ten M}$:
The traceless local field tensor operator (\op{M2L}) is defined as
\begin{equation} \label{eq:field tensor v1}
    \tl{\ten L}_{\mi n}(\vec z' = \vec z + \vec r) = \sum_{|\mi m| \leq p-n} \frac{1}{n!}\frac{m!}{\mi m!} \tl{\ten M}_{\mi m}(\vec z) \vec\nabla^{\mi n + \mi m}\vec r^{-1} 
\end{equation}
From a local field tensor $\tl{\ten L}_{\mi n}(\vec z')$, the final field (or gradient) is approximated at nearby points with the traceless field tensor to particle operator (\op{L2P}):
\begin{equation} \label{eq:phi v1}
    \nabla^{\mi k} \Phi(\vec x = \vec z' + \vec r) \approx \sum_{|\mi n| \leq p-k} 
    \frac{(n+k)!}{k!n!} \frac{n!}{\mi n!}
    \vec r^{\mi n} \tl{\ten L}_{\mi n + \mi k}(\vec z')
\end{equation}
for some $\vec x$ in $B$ centered at $\vec z'$.
Substituting \eqnref{eq:multipole v1} and \eqnref{eq:field tensor v1} into \eqnref{eq:phi v1} simplifies to the original expression \eqnref{eq:PhiAB} for $\Phi_{A\rightarrow B}$.
The power of these operators is clear when one considers that field tensors from multiple non-overlapping regions may be accumulated (i.e., summed element-wise) to form a new field tensor if the centers are the same.
With such a field tensor, \op{L2P} simultaneously applies the effect of multiple regions in a single step.

For the fast multipole method to be fully $O(N)$ two more operators are needed.
Since the choice of multipole expansion center in \op{P2M} was arbitrary this can be changed. 
The operator \op{M2M} recenters an existing traceless multipole about a new point $\vec z'$:
\begin{align} \label{eq:Mshift}
    &\tl{\ten M}_{\mi n}(\vec z' = \vec z - \vec \delta) 
        = \D[\ten M']_{\mi n}\\
    &\ten M'_{\mi n} = \sum_{|\mi k| \leq |\mi n|} \frac{(-1)^{k}}{k!}  
        \left[\frac{k!}{\mi k!} \frac{(n - k)!}{(\mi n - \mi k)!} \frac{\mi n!}{n!} \right]
        \vec \delta^{\mi k} \tl{\ten M}_{\mi n - \mi k}(\vec z)
\end{align}
When two multipole expansions share the same center they may be combined into a single expansion.
The factor in brackets ensures that the direct product between $\vec \delta^{\mi k}$ and $\tl{\ten M}$ produces a symmetric result that can be stored in the compressed tensor format. This is derived from an application of the $\mathrm{Sym}$ operator in \eqnref{eq:Sym}.
Similarly, the traceless field tensor translation operator (\op{L2L}) 
\begin{equation} \label{eq:Fshift}
    \tl{\ten L}_{\mi n}(\vec z' = \vec z + \vec \delta) = \sum_{|\mi k| \leq p-n} \frac{(n+k)!}{n! k!} \frac{k!}{\mi k!}\vec \delta^{\mi k}\tl{\ten L}_{\mi n + \mi k}(\vec z)
\end{equation}
moves the location where a field tensor was evaluated.
As with the multipole expansion, two field tensors sharing the same center may be combined.

\begin{figure}
    \center\includegraphics[width=0.7\columnwidth]{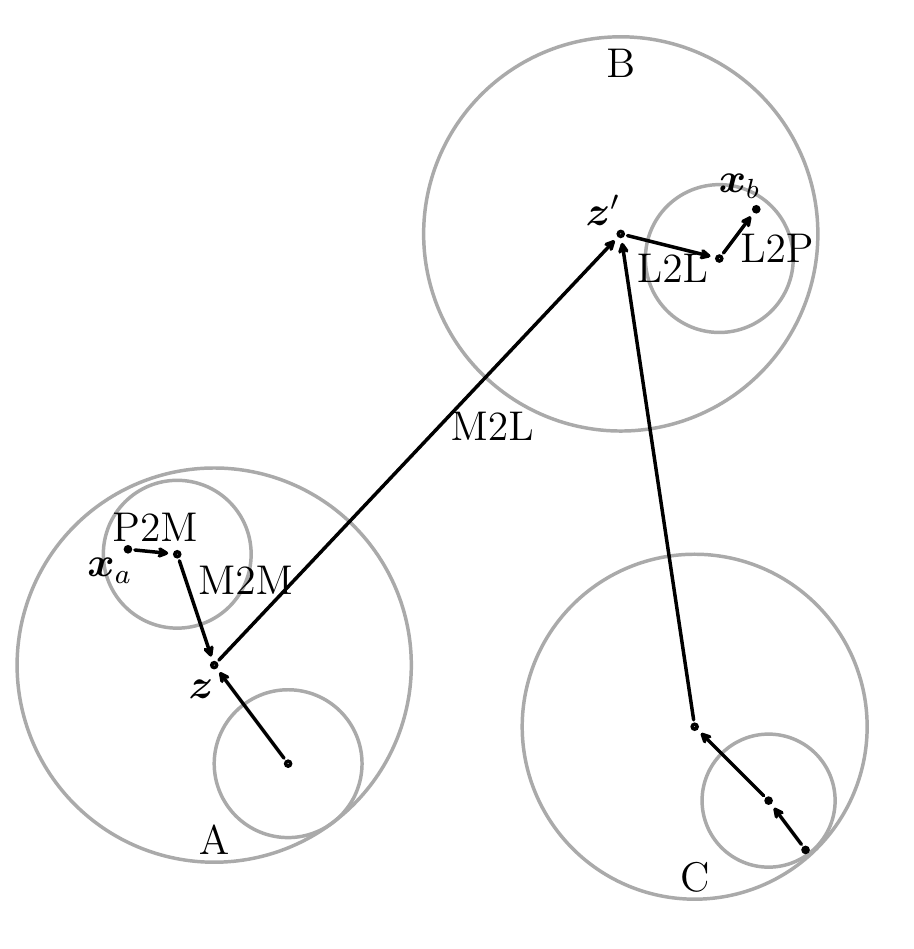}
    \caption{A sketch of the FMM operators. \op{M2M} accumulates multipole expansions at common locations, which allows \op{M2L} to compute cell-cell interactions between large regions. The interaction results in a local field tensor which is shifted with \op{L2L} and finally onto individual particles with \op{L2P} to compute the accumulated potential at a remote point.}

    \label{fig:operators}
\end{figure}

\figref{fig:operators} sketches the hierarchical application of each operator to compute the interaction of two large regions on a third.
Multipole expansions for the particles in the smallest regions are computed using \op{P2M} and combined to larger multipole expansions with \op{M2M}.
Interactions between larger regions are evaluated by computing the local field tensor using \op{M2L}.
This field expansion is then recentered at smaller local regions using \op{L2L} and finally evaluated at particle position with \op{L2P}.
%

\subsection{Dipoles and higher order multipoles\label{sec:Dipoles}}

Some applications in molecular dynamics also include point dipoles (or quadrupoles) $\mathfrak{\vec p}$ in addition to single point charges.
The dipoles capture the polarization effect present in water and many other molecules.
Point dipoles have a net charge of zero, but generate and respond to electric fields as though they were two opposite charged particles infinitely close together.
They can be incorporated into the FMM framework in a straightforward manner by modifying \op{P2M}.
We can construct point dipoles at $\vec r_a$ by shifting with \op{M2M} a pure dipole $\ten M^{\mathrm{dipole}}_{\mi 1} = \mathfrak{\vec p}$, where all other components are zero.
Higher order constructions are similarly possible, but one must be careful that the multipole object is symmetric and traceless before shifting.


\section{The Fast Multipole Method\label{sec:Fast Multipole Method}}
Our FMM implementation is based on the tree code method discussed in \cite{2002JCoPh.179...27D,Yokota:2012:TSF:2403996.2403998}, and \cite{2015JChPh.142b4109C}.
We refer the reader again to a sketch of the procedure in \figref{fig:operators}.

The potential $\Phi^{\mathrm{tot}} = \Phi^{\mathrm{near}} + \Phi^{\mathrm{far}}$ is expressed as near and far components which are handled separately.
The near field is computed using a direct sum $\O(n^2)$ method, where $n \ll N$ is small and does not change with the total system size $N$, so that we can consider it a constant cost.

The volume of $N$ particles is represented hierarchically by a binary tree of nested cells with the smallest (leaf) cells having no more than $N_{\mathrm{bucket}}$ $(\simeq 32)$ particles.
Cells are split across the longest axis such that there are an equal number of particles on each side.
The extents of the child cells are shrunk to tightly bound the enclosed particles. 
\citep[For applications such as in galaxy simulations where the distribution of particles is inhomogeneous, it is recommended to split geometrically at the cell center; for more details see][]{doi:10.1137/S0097539797326307}.
In a perfectly balanced tree there are $N_{\mathrm{cells}} = 2N / N_{\mathrm{bucket}} - 1 \rightarrow \O(N)$ cells.
In each of the smallest cells we calculate a multipole expansion with \op{P2M} \eqnrefp{eq:P2M}.
Each multipole expansion is centered on the geometric center of the cell to minimize the enclosing radius.
Astrophysical applications typically expand around the center of mass to take advantage of the vanishing dipole in this case.
Working from the bottom up, the multipole expansion of a parent cell (with center $\vec z_P$) is computed by recentering each child cell expansion to $z_P$ with \op{M2M} \eqnrefp{eq:M2M} and summing the components.
This is an $\O(N)$ operation for the whole tree.

The interaction phase determines which cell pairs are sufficiently far apart that a multipole interaction is acceptable within a given tolerance.
More distant cells produce a more accurate result but the larger the two regions, the less computation is needed to cover the entire volume.
We use the following multipole acceptance criteria (MAC), with tuning parameter $0 \leq \theta \leq 1$, to decide if two regions are sufficiently well-separated:
If $\theta|\vec z_B - \vec z_A| > |\vec r_A + \vec r_B|$ is satisfied for region centers $\vec z_A,\vec z_B$ with respective enclosing radii $\vec r_A,\vec r_B$ then the mutual interaction is allowed.
For $\theta=0$ no two regions will be accepted and the algorithm will degrade to an $\O(N^2)$ direct summation over the entire volume.

We use the dual-tree algorithm for FMM from \cite{2002JCoPh.179...27D} to find interacting cell pairs.
By its nature, the algorithm ensures that Newton's third law is satisfied and linear momentum is conserved.
Beginning from the root cell, we apply the MAC to the two child cells.
If this is satisfied, we compute the mutual interactions $\tl{\ten L}_{A \rightarrow B}$ and $\tl{\ten L}_{B \rightarrow A}$ using \op{M2L} \eqnrefp{eq:M2L} in both directions.
The resulting field tensors are accumulated at each respective cell (this is implementation specific and it is possible to accumulate this on-the-fly).
If the MAC is not satisfied, the larger cell is opened and each child of the larger cell is compared to the smaller cell.
If a cell cannot be opened because it is a leaf cell, the other cell of the interaction is opened instead.
If two leaf cells do not satisfy the MAC then the interaction is considered part of the near field component and is computed via a direct summation.
Because tree descent is cut off with the MAC and cell-cell interactions are computed, this phase runs in $\O(N)$ time.
It is during this phase that a distributed implementation of FMM may exchange multipole expansion data.
With the traceless tensors, this data exchange is minimized.

The field tensors that have been accumulated and stored throughout the tree are then recursively shifted to the lowest cell level to compute the final field at each particle position.
Starting from the root cell and descending depth-first to the lowest cells, the field tensor at each cell is recentered using \op{L2L} \eqnrefp{eq:L2L} on each of its child cells and accumulated.
This is again an $\O(N)$ process.

The final field tensor at the center of each leaf cell represents the total potential from all other cells.
The potential (or gradients) at each particle position is computed using \op{L2P} \eqnrefp{eq:L2P}.
Since this phase simply loops over all particles it is also an $\O(N)$ process.

In terms of the number of applications of \op{M2M}, the bottom-up construction of the multipole expansions in each tree cell is strictly $O(N_{\mathrm{cells}})$ without any hidden constants.
However, for the interaction phase the number of calls to \op{M2L} is dependent on the geometry of the problem.
In realistic scenarios, \op{M2L} is used far more than \op{M2M}, often by 40--60 times as measured in our experiments, and is therefore the most crucial operator to optimize.


\section{Multi-index Notation\label{sec:mi notation}}
Multi-index notation simplifies the presentation of symmetric tensors.
Tensors (resp. indices) are assumed here to be three dimensional.
An index is written $\mi n = (n_x, n_y, n_z)$, where $n_i \ge 0$.
In the following, $\mi n$ and $\mi m$ are indices with three dimensions, $\vec r$ is a vector, and $k$ a scalar:
$|\mi n|         = n \equiv n_x + n_y + n_z$;
$\mi n!          \equiv n_x!\; n_y!\; n_z!$;
$\mi n \pm k     \equiv (n_x \pm k, n_y \pm k, n_z \pm k)$;
$\mi n \pm \mi m \equiv (n_x \pm m_x, n_y \pm m_y, n_z \pm m_z)$;
$\vec r^{\mi n}  \equiv r_x^{n_x}\, r_y^{n_y}\, r_z^{n_z}$.
Indices  with negative components are forbidden and any expressions containing such indices are to be ignored (equal zero).
The $\mi n$th element of tensor $\ten A$ is the scalar $\ten A_{\mi n}$.
A different notation may be used to refer to specific indices of a three-dimensional tensor, as in $\ten A_{xyyzzz} = \ten A_{(1,2,3)}$ and $\ten A_{xxz} = \ten A_{(2,0,1)}$.


\section{Miscellaneous Details\label{sec:Technical Details}}
The number of elements required for fully traceless and non-traceless tensors is listed in \tabref{tab:m elements}. 
Timing results were obtained on an Intel Sandybridge 2.60 GHz E5-2640v3 CPU.
The multipole generating software was written using Python v2.7.16 and Sympy v1.1.1.
The latest version of the \code{Mosaic} source code is publicly available from: github.com/jpcoles/mosaic
\begin{table}[h]
    {\footnotesize
    \begin{tabular*}{\columnwidth}{@{\extracolsep{\fill}}llllll}
$p$ & \multicolumn{2}{l}{Non-traceless} & \multicolumn{2}{l}{Traceless}  & Savings (\%) \\
\cline{2-3}\cline{4-5} \\
\mbox{} & at $p$th       & total          & at $p$th       & total          & \mbox{} \\
\mbox{} &           order &       elements &           order &       elements & \mbox{} \\
\hline
0 & 1 & 1 & 1 & 1 & 0 \\
1 & 3 & 4 & 3 & 4 & 0 \\
2 & 6 & 10 & 5 & 9 & 10 \\
3 & 10 & 20 & 7 & 16 & 20 \\
4 & 15 & 35 & 9 & 25 & 29 \\
5 & 21 & 56 & 11 & 36 & 36 \\
6 & 28 & 84 & 13 & 49 & 42 \\
\textbf{7} & \textbf{36} & \textbf{120} & \textbf{15} & \textbf{64} & \textbf{47} \\
8 & 45 & 165 & 17 & 81 & 51 \\
9 & 55 & 220 & 19 & 100 & 55 \\
10 & 66 & 286 & 21 & 121 & 58 \\
\end{tabular*}
    }
    \caption{
        Storage size needed for non-traceless and traceless multipole tensors for expansion orders $p \le 10$.
        In our production molecular dynamics runs an expansion order $p=7$ is used.
        The last column shows by how much the storage is reduced when using the traceless operator. This has important consequences for memory storage and network communication.
    }

    \label{tab:m elements}
\end{table}

\bibliography{ms}

\begin{thebibliography}{24}
\expandafter\ifx\csname natexlab\endcsname\relax\def\natexlab#1{#1}\fi
\providecommand{\url}[1]{\texttt{#1}}
\providecommand{\href}[2]{#2}
\providecommand{\path}[1]{#1}
\providecommand{\DOIprefix}{doi:}
\providecommand{\ArXivprefix}{arXiv:}
\providecommand{\URLprefix}{URL: }
\providecommand{\Pubmedprefix}{pmid:}
\providecommand{\doi}[1]{\href{http://dx.doi.org/#1}{\path{#1}}}
\providecommand{\Pubmed}[1]{\href{pmid:#1}{\path{#1}}}
\providecommand{\bibinfo}[2]{#2}
\ifx\xfnm\relax \def\xfnm[#1]{\unskip,\space#1}\fi
\bibitem[{Anderson(1999)}]{doi:10.1137/S0097539797326307}
\bibinfo{author}{Anderson, R.}, \bibinfo{year}{1999}.
\newblock \bibinfo{title}{Tree data structures for n-body simulation}.
\newblock \bibinfo{journal}{SIAM Journal on Computing} \bibinfo{volume}{28},
  \bibinfo{pages}{1923--1940}.
\bibitem[{Applequist(1984)}]{APPLEQUIST1984279}
\bibinfo{author}{Applequist, J.}, \bibinfo{year}{1984}.
\newblock \bibinfo{title}{Fundamental relationships in the theory of electric
  multipole moments and multipole polarizabilities in static fields}.
\newblock \bibinfo{journal}{Chemical Physics} \bibinfo{volume}{85},
  \bibinfo{pages}{279 -- 290}.
\bibitem[{Applequist(1989)}]{Applequist1989}
\bibinfo{author}{Applequist, J.}, \bibinfo{year}{1989}.
\newblock \bibinfo{title}{Traceless cartesian tensor forms for spherical
  harmonic functions: new theorems and applications to electrostatics of
  dielectric media}.
\newblock \bibinfo{journal}{Journal of Physics A: Mathematical and General}
  \bibinfo{volume}{22}, \bibinfo{pages}{4303}.
\bibitem[{{Barnes} and {Hut}(1986)}]{1986Natur.324..446B}
\bibinfo{author}{{Barnes}, J.}, \bibinfo{author}{{Hut}, P.},
  \bibinfo{year}{1986}.
\newblock \bibinfo{title}{{A hierarchical O(N log N) force-calculation
  algorithm}}.
\newblock \bibinfo{journal}{\nat} \bibinfo{volume}{324},
  \bibinfo{pages}{446--449}.
\bibitem[{Burgos and Bonadeo(1981)}]{doi:10.1080/00268978100102251}
\bibinfo{author}{Burgos, E.}, \bibinfo{author}{Bonadeo, H.},
  \bibinfo{year}{1981}.
\newblock \bibinfo{title}{Electrical multipoles and multipole interactions:
  compact expressions and a diagrammatic method}.
\newblock \bibinfo{journal}{Molecular Physics} \bibinfo{volume}{44},
  \bibinfo{pages}{1--15}.
\bibitem[{{Coles} and {Masella}(2015)}]{2015JChPh.142b4109C}
\bibinfo{author}{{Coles}, J.P.}, \bibinfo{author}{{Masella}, M.},
  \bibinfo{year}{2015}.
\newblock \bibinfo{title}{{The fast multipole method and point dipole moment
  polarizable force fields}}.
\newblock \bibinfo{journal}{{J. Chem. Phys.}} \bibinfo{volume}{142},
  \bibinfo{pages}{024109}.
\bibitem[{{Darden} et~al.(1993){Darden}, {York} and
  {Pedersen}}]{1993JChPh..9810089D}
\bibinfo{author}{{Darden}, T.}, \bibinfo{author}{{York}, D.},
  \bibinfo{author}{{Pedersen}, L.}, \bibinfo{year}{1993}.
\newblock \bibinfo{title}{{Particle mesh Ewald: An N . log(N) method for Ewald
  sums in large systems}}.
\newblock \bibinfo{journal}{\jcp} \bibinfo{volume}{98},
  \bibinfo{pages}{10089--10092}.
\bibitem[{{Dehnen}(2002)}]{2002JCoPh.179...27D}
\bibinfo{author}{{Dehnen}, W.}, \bibinfo{year}{2002}.
\newblock \bibinfo{title}{{A Hierarchical $O(N)$ Force Calculation Algorithm}}.
\newblock \bibinfo{journal}{Journal of Computational Physics}
  \bibinfo{volume}{179}, \bibinfo{pages}{27--42}.
\bibitem[{{Dehnen}(2014)}]{2014ComAC...1....1D}
\bibinfo{author}{{Dehnen}, W.}, \bibinfo{year}{2014}.
\newblock \bibinfo{title}{{A fast multipole method for stellar dynamics}}.
\newblock \bibinfo{journal}{Computational Astrophysics and Cosmology}
  \bibinfo{volume}{1}, \bibinfo{pages}{1}.
\bibitem[{Greengard et~al.(1998)Greengard, Huang, Rokhlin and
  Wandzura}]{Greengard:1998:AFM:615257.615572}
\bibinfo{author}{Greengard, L.}, \bibinfo{author}{Huang, J.},
  \bibinfo{author}{Rokhlin, V.}, \bibinfo{author}{Wandzura, S.},
  \bibinfo{year}{1998}.
\newblock \bibinfo{title}{Accelerating fast multipole methods for the helmholtz
  equation at low frequencies}.
\newblock \bibinfo{journal}{IEEE Comput. Sci. Eng.} \bibinfo{volume}{5},
  \bibinfo{pages}{32--38}.
\newblock \DOIprefix\doi{10.1109/99.714591}.
\bibitem[{{Greengard} and {Rokhlin}(1987)}]{1987JCoPh..73..325G}
\bibinfo{author}{{Greengard}, L.}, \bibinfo{author}{{Rokhlin}, V.},
  \bibinfo{year}{1987}.
\newblock \bibinfo{title}{{A fast algorithm for particle simulations}}.
\newblock \bibinfo{journal}{Journal of Computational Physics}
  \bibinfo{volume}{73}, \bibinfo{pages}{325--348}.
\bibitem[{Greengard and Rokhlin(1997)}]{greengard_rokhlin_1997}
\bibinfo{author}{Greengard, L.}, \bibinfo{author}{Rokhlin, V.},
  \bibinfo{year}{1997}.
\newblock \bibinfo{title}{A new version of the fast multipole method for the
  laplace equation in three dimensions}.
\newblock \bibinfo{journal}{Acta Numerica} \bibinfo{volume}{6},
  \bibinfo{pages}{229–269}.
\bibitem[{Huang et~al.(2018)Huang, Luo, Li, Chen and Zhang}]{HUANG2018122}
\bibinfo{author}{Huang, H.}, \bibinfo{author}{Luo, L.S.}, \bibinfo{author}{Li,
  R.}, \bibinfo{author}{Chen, J.}, \bibinfo{author}{Zhang, H.},
  \bibinfo{year}{2018}.
\newblock \bibinfo{title}{Improve the efficiency of the cartesian tensor based
  fast multipole method for coulomb interaction using the traces}.
\newblock \bibinfo{journal}{Journal of Computational Physics}
  \bibinfo{volume}{371}, \bibinfo{pages}{122 -- 136}.
\bibitem[{Lagard{\`e}re et~al.(2018)Lagard{\`e}re, Jolly, Lipparini, Aviat,
  Stamm, Jing, Harger, Torabifard, Cisneros, Schnieders, Gresh, Maday, Ren,
  Ponder and Piquemal}]{C7SC04531J}
\bibinfo{author}{Lagard{\`e}re, L.}, \bibinfo{author}{Jolly, L.H.},
  \bibinfo{author}{Lipparini, F.}, \bibinfo{author}{Aviat, F.},
  \bibinfo{author}{Stamm, B.}, \bibinfo{author}{Jing, Z.F.},
  \bibinfo{author}{Harger, M.}, \bibinfo{author}{Torabifard, H.},
  \bibinfo{author}{Cisneros, G.A.}, \bibinfo{author}{Schnieders, M.J.},
  \bibinfo{author}{Gresh, N.}, \bibinfo{author}{Maday, Y.},
  \bibinfo{author}{Ren, P.Y.}, \bibinfo{author}{Ponder, J.W.},
  \bibinfo{author}{Piquemal, J.P.}, \bibinfo{year}{2018}.
\newblock \bibinfo{title}{Tinker-hp: a massively parallel molecular dynamics
  package for multiscale simulations of large complex systems with advanced
  point dipole polarizable force fields}.
\newblock \bibinfo{journal}{Chem. Sci.} \bibinfo{volume}{9},
  \bibinfo{pages}{956--972}.
\bibitem[{Lorenzen et~al.(2012)Lorenzen, Schw{\"o}rer, Tr{\"o}ster, Mates and
  Tavan}]{Lorenzen:2012aa}
\bibinfo{author}{Lorenzen, K.}, \bibinfo{author}{Schw{\"o}rer, M.},
  \bibinfo{author}{Tr{\"o}ster, P.}, \bibinfo{author}{Mates, S.},
  \bibinfo{author}{Tavan, P.}, \bibinfo{year}{2012}.
\newblock \bibinfo{title}{Optimizing the accuracy and efficiency of fast
  hierarchical multipole expansions for md simulations}.
\newblock \bibinfo{journal}{Journal of Chemical Theory and Computation}
  \bibinfo{volume}{8}, \bibinfo{pages}{3628--3636}.
\bibitem[{Masella et~al.(2008)Masella, Borgis and
  Cuniasse}]{doi:10.1002/jcc.20932}
\bibinfo{author}{Masella, M.}, \bibinfo{author}{Borgis, D.},
  \bibinfo{author}{Cuniasse, P.}, \bibinfo{year}{2008}.
\newblock \bibinfo{title}{Combining a polarizable force-field and a
  coarse-grained polarizable solvent model: Application to long dynamics
  simulations of bovine pancreatic trypsin inhibitor}.
\newblock \bibinfo{journal}{Journal of Computational Chemistry}
  \bibinfo{volume}{29}, \bibinfo{pages}{1707--1724}.
\bibitem[{Masella et~al.(2011)Masella, Borgis and
  Cuniasse}]{doi:10.1002/jcc.21846}
\bibinfo{author}{Masella, M.}, \bibinfo{author}{Borgis, D.},
  \bibinfo{author}{Cuniasse, P.}, \bibinfo{year}{2011}.
\newblock \bibinfo{title}{Combining a polarizable force-field and a
  coarse-grained polarizable solvent model. ii. accounting for hydrophobic
  effects}.
\newblock \bibinfo{journal}{Journal of Computational Chemistry}
  \bibinfo{volume}{32}, \bibinfo{pages}{2664--2678}.
\bibitem[{Masella et~al.(2013)Masella, Borgis and
  Cuniasse}]{doi:10.1002/jcc.23237}
\bibinfo{author}{Masella, M.}, \bibinfo{author}{Borgis, D.},
  \bibinfo{author}{Cuniasse, P.}, \bibinfo{year}{2013}.
\newblock \bibinfo{title}{A multiscale coarse-grained polarizable solvent model
  for handling long tail bulk electrostatics}.
\newblock \bibinfo{journal}{Journal of Computational Chemistry}
  \bibinfo{volume}{34}, \bibinfo{pages}{1112--1124}.
\bibitem[{Meurer et~al.(2017)Meurer, Smith, Paprocki, \v{C}ert\'{i}k,
  Kirpichev, Rocklin, Kumar, Ivanov, Moore, Singh, Rathnayake, Vig, Granger,
  Muller, Bonazzi, Gupta, Vats, Johansson, Pedregosa, Curry, Terrel,
  Rou\v{c}ka, Saboo, Fernando, Kulal, Cimrman and
  Scopatz}]{10.7717/peerj-cs.103}
\bibinfo{author}{Meurer, A.}, \bibinfo{author}{Smith, C.P.},
  \bibinfo{author}{Paprocki, M.}, \bibinfo{author}{\v{C}ert\'{i}k, O.},
  \bibinfo{author}{Kirpichev, S.B.}, \bibinfo{author}{Rocklin, M.},
  \bibinfo{author}{Kumar, A.}, \bibinfo{author}{Ivanov, S.},
  \bibinfo{author}{Moore, J.K.}, \bibinfo{author}{Singh, S.},
  \bibinfo{author}{Rathnayake, T.}, \bibinfo{author}{Vig, S.},
  \bibinfo{author}{Granger, B.E.}, \bibinfo{author}{Muller, R.P.},
  \bibinfo{author}{Bonazzi, F.}, \bibinfo{author}{Gupta, H.},
  \bibinfo{author}{Vats, S.}, \bibinfo{author}{Johansson, F.},
  \bibinfo{author}{Pedregosa, F.}, \bibinfo{author}{Curry, M.J.},
  \bibinfo{author}{Terrel, A.R.}, \bibinfo{author}{Rou\v{c}ka, v.},
  \bibinfo{author}{Saboo, A.}, \bibinfo{author}{Fernando, I.},
  \bibinfo{author}{Kulal, S.}, \bibinfo{author}{Cimrman, R.},
  \bibinfo{author}{Scopatz, A.}, \bibinfo{year}{2017}.
\newblock \bibinfo{title}{Sympy: symbolic computing in python}.
\newblock \bibinfo{journal}{PeerJ Computer Science} \bibinfo{volume}{3},
  \bibinfo{pages}{e103}.
\bibitem[{{Potter} et~al.(2017){Potter}, {Stadel} and
  {Teyssier}}]{2017ComAC...4....2P}
\bibinfo{author}{{Potter}, D.}, \bibinfo{author}{{Stadel}, J.},
  \bibinfo{author}{{Teyssier}, R.}, \bibinfo{year}{2017}.
\newblock \bibinfo{title}{{PKDGRAV3: beyond trillion particle cosmological
  simulations for the next era of galaxy surveys}}.
\newblock \bibinfo{journal}{Computational Astrophysics and Cosmology}
  \bibinfo{volume}{4}, \bibinfo{pages}{2}.
\bibitem[{Shanker and Huang(2007)}]{SHANKER2007732}
\bibinfo{author}{Shanker, B.}, \bibinfo{author}{Huang, H.},
  \bibinfo{year}{2007}.
\newblock \bibinfo{title}{Accelerated cartesian expansions -- a fast method for
  computing of potentials of the form r−ν for all real ν}.
\newblock \bibinfo{journal}{Journal of Computational Physics}
  \bibinfo{volume}{226}, \bibinfo{pages}{732 -- 753}.
\bibitem[{{Stadel}(2001)}]{2001PhDT........21S}
\bibinfo{author}{{Stadel}, J.G.}, \bibinfo{year}{2001}.
\newblock \bibinfo{title}{{Cosmological N-body simulations and their
  analysis}}.
\newblock Ph.D. thesis. University of Washington.
\bibitem[{Yokota(2013)}]{doi:10.1260/1748-3018.7.3.301}
\bibinfo{author}{Yokota, R.}, \bibinfo{year}{2013}.
\newblock \bibinfo{title}{An fmm based on dual tree traversal for many-core
  architectures}.
\newblock \bibinfo{journal}{Journal of Algorithms \& Computational Technology}
  \bibinfo{volume}{7}, \bibinfo{pages}{301--324}.
\bibitem[{Yokota and Barba(2012)}]{Yokota:2012:TSF:2403996.2403998}
\bibinfo{author}{Yokota, R.}, \bibinfo{author}{Barba, L.A.},
  \bibinfo{year}{2012}.
\newblock \bibinfo{title}{A tuned and scalable fast multipole method as a
  preeminent algorithm for exascale systems}.
\newblock \bibinfo{journal}{Int. J. High Perform. Comput. Appl.}
  \bibinfo{volume}{26}, \bibinfo{pages}{337--346}.

\end{thebibliography}

\end{document}